\documentclass[letterpaper, 10 pt, conference]{ieeeconf}  

\usepackage{color,soul}
\IEEEoverridecommandlockouts                              

\usepackage{amsmath} 
\usepackage{amssymb}  
\usepackage{cite}
\usepackage{graphicx}
\usepackage{setspace}
\usepackage{mathrsfs}
\usepackage[ruled,vlined,titlenumbered]{algorithm2e}
\usepackage{balance}
\usepackage{multirow}
\usepackage{multicol}
\usepackage{makecell}
\usepackage{array, caption, tabularx,  ragged2e,  booktabs}
\usepackage{CJKutf8}
\usepackage{url}

\newlength\mylen

\title{\LARGE \bf
Generating Socially Acceptable Perturbations for Efficient Evaluation of Autonomous Vehicles}

\author{Songan Zhang$^{1}$, Huei Peng$^{2}$, Subramanya Nageshrao$^{3}$ and H. Eric Tseng$^{3}$
\thanks{*This work was supported by Ford Motor Company}
\thanks{$^{1}$Songan Zhang is a Ph.D. Student in Mechanical Engineering, University of Michigan, email:
        {\tt\small songanz@umich.edu}}%
\thanks{$^{2}$Huei Peng is a professor of Mechanical Engineering, University of Michigan, email:
        {\tt\small hpeng@umich.edu}}%
\thanks{$^{3}$ Subramanya Nageshrao and H. Eric Tseng are research scientists in the Ford Motor Company.}%
}

\begin{document}

\maketitle
\thispagestyle{empty}
\pagestyle{empty}

\begin{abstract}

Deep reinforcement learning methods have been widely used in recent years for autonomous vehicle's decision-making. A key issue is that deep neural networks can be fragile to adversarial attacks or other unseen inputs. In this paper, we address the latter issue: we focus on generating socially acceptable perturbations (SAP), so that the autonomous vehicle (AV agent), instead of the challenging vehicle (attacker), is primarily responsible for the crash. In our process, one attacker is added to the environment and trained by deep reinforcement learning to generate the desired perturbation. The reward is designed so that the attacker aims to fail the AV agent in a socially acceptable way. After training the attacker, the agent policy is evaluated in both the original naturalistic environment and the environment with one attacker. The results show that the agent policy which is safe in the naturalistic environment has many crashes in the perturbed environment.

\end{abstract}

\section{INTRODUCTION}
\label{INTRODUCTION}
Over the past two decades, deep learning algorithms (DLA) and deep natural network (DNN) have been widely-used in different fields. From image recognition \cite{krizhevsky2012imagenet}, speech recognition \cite{hannun2014deep}, to natural language processing \cite{sutskever2014sequence}.  However, recent research shows that DNNs may be vulnerable to adversarial perturbations. In \cite{brown2017adversarial}, the authors found that adversarial image patches can lead white-box DNNs to erroneous classification results. Papernot et al. in \cite{papernot2017practical} further developed an attack using synthetic data generation, to craft adversarial image examples mis-classified by black-box DNNs. A more comprehensive overview of adversarial attack on DNN can be found in \cite{balda2020adversarial}.

DNNs have been used in the field of deep reinforcement learning (DRL), where the goal is to train an agent to maximize the expected return. DNN can work as an actor net or a critic net, to provide the optimal policy directly or estimate the expected future return for different actions. DRL policies are vulnerable to adversarial perturbations. In \cite{xiao2019characterizing}, the authors characterize different types of attack on DRL, which can be attacked by adding perturbation to observations or environment transition probabilities. To perturb observations, researchers first follows the same ideas as attacking DNN, which leads the DRL policy to a different action \cite{huang2017adversarial, lin2017tactics}.  In \cite{xiao2019characterizing}, the environment transition model is perturbed. However, as pointed out by the authors, these attacks are useful only under very specific conditions. 

In the field of autonomous vehicles (AV), researchers also tried to attack existing AV system for the purpose of evaluation or synthesis. In a recent report \cite{Tencent2019Experimental}, Tencent’s Keen Security Lab showed how they were able to bamboozle a Tesla Model S into switching lanes to drive directly into the oncoming traffic by manipulating the input video. This attack is in the category of observation attack. To our best knowledge, there has been little study on attacking the environment transition model, which is the main contribution of this paper. 

This paper will focus on finding an adversarial policy (i.e. for the attacker) to socially acceptable attacks for a given victim agent. In other words, we focus on manipulating the environment transition function (by adding an attacker) which is limited to attacks that when crashes are generated, the responsibility is attributed to the AV agent. The paper is organized as follows. In Section \ref{RELATED WORKS}, related works are introduced, followed by Section \ref{PRELIMINARIES}, where we give a brief review on reinforcement learning. Section \ref{THE AV POLICY} describes the details of the victim policy. In Section \ref{SOCIALLY ACCEPTABL ATTACKS}, we introduce the training environment and develop socially acceptable attack design. The simulation setup and results are shown in Section \ref{SIMULATION SETUP} and \ref{RESULTS}, respectively. Our paper is concluded by Section \ref{CONCLUSIONS}.

\section{RELATED WORKS}
\label{RELATED WORKS}

One of the most widely used adversarial attack techniques is the fast gradient sign method (FGSM). It is originally used in image recognition attack.  With small modification, policies obtained through reinforcement learning can also be attacked by this method. With the assumption of white-box attacks, in \cite{huang2017adversarial}, the authors use FGSM to generate adversarial observations, leading to a pre-trained policy to lose the Pong game. Lin et al. in \cite{lin2017tactics} further developed an enchanting attack aimed at maliciously luring an agent to a certain state. Xiao et al. in \cite{xiao2019characterizing} extended FGSM to black-box attacks via imitation learning and other methods.

In general, an attacker does not have direct access to the victim's observations. Under this assumption, Gleave et al. \cite{gleave2019adversarial} demonstrate the existence of adversarial policies in zero-sum games between humanoid robots against black-box victims trained via state-of-the-art reinforcement learning. The adversarial policies reliably win against the victims but generate seemingly random and uncoordinated behavior which is definitely not a ``direct'' attack. Although using multi-agent reinforcement learning to solve a zero-sum Markov game has a long history (from 1994 \cite{littman1994markov}), \cite{gleave2019adversarial} inspired us to think: can we design an attacker car which can cause failures of the victim car (crash) without directly crash into it? 

We managed to achieve this objective based on results presented in other related works. In \cite{shalev2017formal}, the authors introduced a responsibility-sensitive safety (RSS) model which formalizes an interpretation of “Duty of Care” from Tort law. The Duty of Care states that an individual should exercise “reasonable care” while performing acts that have the potential harm others. RSS is a rigorous mathematical model formalizing an interpretation of the law which is applicable to self-driving cars. This RSS mathematical model has been successfully implemented in NHTSA pre-crash situation in \cite{Shashua2018}. By interpreting the responsibility, we can attack the victim in the way described in \cite{xiao2019characterizing}, which perturbs the environment transition model, with the requirement that reward is given only when the victim is responsible for the crash. 

In this paper, we use the Markov game framework since there is an attacker and a victim in this environment. Finding an adversarial policy is a single-agent reinforcement learning problem because the victim's policy is fixed. 

\section{REINFORCEMENT LEARNING BASICS}
\label{PRELIMINARIES}

In reinforcement learning, an agent is trying to learn an optimal policy to maximize the cumulative reward interacting with the environment. Usually the problem is modeled as a Markov decision process (MDP), defined as: $\mathcal{M = (S,A,P},R,\gamma)$, where $\mathcal{S}\subseteq \mathbb{R}^n$ is the state, $\mathcal{A} \subseteq \mathbb{R}^m$ is the action, and $\mathcal{P:S\times A \rightarrow S}$ is the transition dynamics which is usually stochastic. And $R:\mathcal{S\times A \rightarrow} \mathbb{R}$ is the reward function while $\gamma \in [0,1)$ is the discount factor. 

For each time step $t$, the agent tries to learn a policy $\pi_\alpha(s_t)=a_t$ with parameters $\alpha$, where $s_t \in \mathcal{S}$ is the state at time $t$  and $a_t \in \mathcal{A}$ is the action at time $t$. The expectation of future cumulative reward starting from state $s$ following policy $\pi_\alpha$ can be described as: 
\begin{equation}
    V(s|\pi_\alpha) = E_{\pi_\alpha,\mathcal{M}}\left[\sum^{\infty}_{t=0} \gamma^t r_t \Bigr\vert s_0=s\right],
\end{equation}
where $r_t$ is the reward at time $t$ and $V$ is the value function. And Q function (action value function) is defined as:
\begin{equation}
    Q(s,a|\pi_\alpha) = E_{\pi_\alpha,\mathcal{M}}\left[\sum^{\infty}_{t=0} \gamma^t r_t \Bigr\vert s_0=s, a_0 = a\right].
\end{equation}

The goal of reinforcement learning is to maximize the expected accumulative reward, i.e.  $E_\pi\left[\sum^{\infty}_{t=0} \gamma^t r_t \right]$. For a discrete action space, the optimal policy can be obtained by learning the accurate Q function $Q^*$ following the optimal policy $\pi_{opt}$ and thus we have $\pi_{opt}(s) = \arg\max_a(Q^*(s,a|\pi_{opt}))$ which is the greedy policy. Accurate Q function with the optimal policy must satisfy the Bellman equation:
\begin{equation}
\label{eq:Bell}
    Q^*(s,a|\pi_{opt}) = r + \gamma E_{s'}\left[\max_{a'}Q^*(s',a')\right],
\end{equation}
where $s'$ is the next state. 

For high dimensional state space or continuous state space, functional approximation of the Q value is necessary to ensure tractability of the solution. While Q-learning for continuous state space with approximation could cause the Q-network to diverge \cite{tsitsiklis1997analysis}, Deep Q-Network (DQN) \cite{mnih2013playing} has successfully demonstrated value function convergence with empirical results using experience replay buffer and target network.

In this work, we use Double Deep Q-Network (DDQN) as the reinforcement learning algorithm. DDQN is a variation of DQN which uses a target Q-Network to select actions for the evaluation of the next Q value. The technique addresses the problem of overestimating future return (i.e. overoptimism). For details, please refer to the original paper \cite{van2016deep}.

\section{THE AV POLICY}
\label{THE AV POLICY}
In this section, the environment for training the victim AV agent is described. We study a discretionary lane change decision making problem in this paper. The state space, action space, the victim training reward, and the simulation environment are introduced. For benchmark purposes, the problem definition and the simulation environment are the same as the one used in \cite{nageshrao2019autonomous}.

\subsection{Training Environment}
\label{Training Environment}
\begin{figure}[h]
    \centering
    \includegraphics[width=0.48\textwidth]{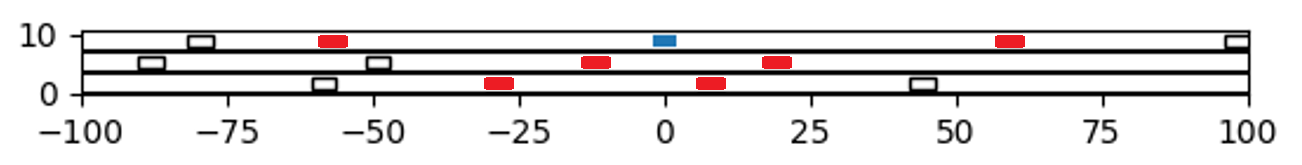}
    \caption{Three lane highway simulator. The blue box: the AV; red boxes: 6 nearest surrounding vehicles; empty boxes: unobserved surrounding vehicles}
    \label{fig:Simulator}
\end{figure}

The victim environment used in this work is a three lane highway simulator based on \cite{nageshrao2019autonomous}. The AV is driving with up to six surrounding vehicles (three vehicles in front, three vehicles behind) as shown in Fig. \ref{fig:Simulator}. The blue box is the AV and the 6 red boxes are the six surrounding vehicles whose states are observed. The remaining boxes are environment vehicles whose states are not observed. The surrounding vehicles driving strategy is also descried in \cite{nageshrao2019autonomous}.

The state space $\mathcal{S}\subseteq\mathbb{R}^n$ of the learning agent (AV) includes AV's lateral position $y$, longitudinal velocity $v_x$ and the relative longitudinal position of the $i^{\mathrm{th}}$ surrounding vehicle $\Delta x^i$, and the relative lateral position of the $i^{\mathrm{th}}$ surrounding vehicle $\Delta y^i$ and the relative longitudinal velocity of the $i^{\mathrm{th}}$ surrounding vehicle $\Delta v_x^i$. In total, we have a state space of $2 + 3\times6\mathrm{(cars)}=20$ dimensions, i.e. $\mathcal{S}\subseteq\mathbb{R}^{20}$.

The actions of both the AV and the surrounding vehicles are discrete. As defined in \cite{nageshrao2019autonomous}, we consider four action choices along the longitudinal direction $a_x$, namely, \textit{maintain speed}, \textit{accelerate}, \textit{brake}, and \textit{hard brake}. Whereas for lateral actions $a_y$, we assume three choices, \textit{lane keep}, \textit{change lane to right}, and \textit{change lane to left}. In total, we define 12 different discrete actions $a=[a_x, a_y]$. For detailed action parameters, please refer to \cite{nageshrao2019autonomous}.

\subsection{Reward Function}

The reward $r^v$ is as defined in \cite{nageshrao2019autonomous}. It is formulated as a function of $(dx, y, v_x)$, where $dx$ is the distance between the AV and its lead vehicle, $y$ is the lateral position of the AV and $v_x$ is its longitudinal velocity. The reward is defined as:
\begin{equation}
    r_x = 
    \begin{cases}
      \exp{\left(-\frac{(dx-dx_{safe})^2}{10dx_{safe}}\right)}-1, & \text{if}\ dx<dx_{safe} \\
      0, & \text{otherwise}
    \end{cases}
\end{equation}

\begin{equation}
    r_y = \exp{\left(-\frac{(y-y_{des})^2}{y_{norm}}\right)}-1,
\end{equation}

\begin{equation}
    r_v = \exp{\left(-\frac{(v_x-v_{des})^2}{v_{norm}}\right)}-1,
\end{equation}
where $dx_{safe}$, $y_{des}$ and $v_{des}$ are the safe longitudinal distance to the lead vehicle, the target lane position, and desired speed, respectively. These three rewards are normalized by $10dx_{safe}$, $y_{norm}$ and $v_{norm}$, respectively, so that no single reward dominates the total reward. Then we have $r^v = \frac{1}{3}(r_x + r_y + r_v)$ if no collision and $r^e = -2$ if collision occurs. 

After learning in an environment without attacker, the AV learned to keep a safe distance from the front vehicle, drive in the center of a lane and drive as fast as possible without violating the speed limit. On top of that, we implemented a ``short-horizon safety check'' that evaluates the action chosen by the agent based on traffic rules and provides an alternative action when needed \cite{nageshrao2019autonomous}.  


\section{SOCIALLY ACCEPTABLE ATTACKS}
\label{SOCIALLY ACCEPTABL ATTACKS}

In this Section, the process for obtaining socially acceptable attacks will be introduced. As discussed in Section \ref{INTRODUCTION} and \ref{RELATED WORKS}, we want to attack the AV agent by perturbing the environment transition probability with the requirement that the AV agent takes the responsibility of the crash. To do that, we add an attack agent near the AV agent and train it with reward considering socially acceptable attacks.

\subsection{Attacker Training Environment}

The state space includes information from 6 surrounding vehicles and the AV. As defined in \ref{Training Environment}, the information includes each vehicle's relative longitudinal position, relative lateral position and relative speed. In addition, we also include the AV's action in the attacker's state space. Therefore, in total we have a 24-dimension state space, i.e. $\mathcal{S}\subseteq\mathbb{R}^{24}$. All the states are scaled for neural network training.  

\begin{figure}[h]
    \centering
    \includegraphics[width=0.48\textwidth]{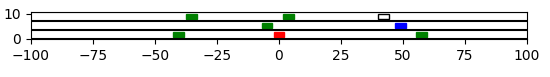}
    \caption{Attacker's training environment. The blue box: the AV; the red box: the attacker; green boxes: other surrounding vehicles in attacker's state space; empty boxes: unobserved surrounding vehicles}
    \label{fig:SimulatorAtt}
\end{figure}

At initialization, the attacker is located near the AV. As shown in Fig. \ref{fig:SimulatorAtt}, the attacker (the red box) can be observed by the AV (the blue box). Although the attacker can observe AV's relative position, velocity and its action, the attacker do not have direct access to the AV's policy. Therefore, we are preforming a black-box attack. 

\subsection{Reward Considering Socially Acceptable Attack}
\label{ATTACK REWARD}

The key component for training a socially acceptable attacker's policy is the reward. The attack agent is rewarded if it causes a collision between the AV and one of the environment cars (not necessarily the attacker), in which the AV is at fault per RSS. In this work, we encode the RSS \cite{shalev2017formal} model and traffic rules through associated reward to train the attacker. Here we recall the 5 ``common sense'' rules followed by RSS \cite{shalev2017formal}:
\begin{enumerate}
    \item Do not hit someone from behind.
    \item Do not cut-in recklessly.
    \item Right-of-way is given, not taken.
    \item Be careful of areas with limited visibility.
    \item If you can avoid an accident without causing another one, you must do it.
\end{enumerate}

\begin{table*}[]
\captionof{table}{Reward design for socially acceptable attack}
\label{table:attackerReward}
\centering
\begin{tabularx}{0.99\textwidth}{c|c|c|c|c|c|c|c}
    \multirow{2}{*}{Pre-crash situation} & \multirow{2}{*}{The Responsible Car} & \multicolumn{2}{c|}{Minimal Evasive Effort (MEE)} & \multicolumn{3}{c|}{Attacker's reward} & \multirow{2}{*}{\parbox{0.4cm}{FC}} \\\cline{3-7}
    & & Responsible car & Irresponsible car & Fault & MEE & Reward & \\\hline
    \multirow{4}{*}{\parbox{3cm}{No car is on the lane marker}} & \multirow{4}{*}{The rear car} & \multirow{4}{*}{Hard brake} & \multirow{4}{*}{-} & \multirow{2}{*}{Env. car} & No & -1 & 0 \\\cline{6-8}
    & & & & & Yes & -0.5 & 1 \\\cline{5-8}
    & & & & \multirow{2}{*}{Victim} & No & 1 & 2 \\\cline{6-8}
    & & & & & Yes & 0.5 & 3 \\\hline
    \hline
    
    \multicolumn{8}{l}{Only one car is on the lane marker and crash with the car in the original lane: Same as no car is on the lane marker} \\\hline
      
    \multirow{4}{*}{\parbox{3cm}{Only one car is on the lane marker:\\ crash with the car in the target lane}} & \multirow{4}{*}{The lane-change car} & \multirow{4}{*}{\parbox{2cm}{Abandon the lane change}} & \multirow{4}{*}{-} & \multirow{2}{*}{Env. car} & No & -1 & 0 \\\cline{6-8}
    & & & & & Yes & -0.5 & 1 \\\cline{5-8}
    & & & & \multirow{2}{*}{Victim} & No & 1 & 4 \\\cline{6-8}
    & & & & & Yes & 0.5 & 5 \\\hline
    \hline
    \multicolumn{8}{l}{Both cars are on the same lane marker: Same as no car is on the lane marker} \\\hline
      
    \multirow{4}{*}{\parbox{3cm}{Both cars are on the lane marker:\\ different lane markers}} & \multirow{4}{*}{\parbox{3cm}{Share responsibility, but left car is the principal responsible car}} & \multirow{4}{*}{\parbox{2cm}{Abandon the lane change}} & \multirow{4}{*}{-} & \multirow{2}{*}{Env. car} & No & -0.8 & 0 \\\cline{6-8}
    & & & & & Yes & -0.3 & 1 \\\cline{5-8}
    & & & & \multirow{2}{*}{Victim} & No & 0.8 & 6 \\\cline{6-8}
    & & & & & Yes & 0.3 & 7 \\\hline
      
\end{tabularx}
\end{table*}

The first three ``common sense'' principles above are related to traffic rules and implemented through associated reward. The forth is not applicable in our highway scenario. To implement the fifth, we refer to another MOBILEYE paper \cite{Shashua2018}. The MOBILEYE group has implemented the RSS model on NHTSA pre-crash scenarios in \cite{Shashua2018}, where they define ``proper response'' to dangerous situations as using Minimal Evasive Effort (MEE). MEE deals with cases in which extra caution is applied to prevent potential situations in which responsibility might be shared. Here we extend the MEE concept, and try to create situations where only one car (the AV) is responsible for the crash. We still expect both cars involved in the crash to use MEE. For instance, in the rear-end collision (which is the rear car's fault), we still expect the rear car to brake before crash. In this work, we only define the MEE for the responsible car for simplicity, but the MEE for the irresponsible car can be defined similarly. Also, in this paper, the action space is discrete. However, this concept can be extended to continuous action space. 

In this work, the reward function is $r(s, a)$, where the $s$ is the state, $a$ is the attacker's action at state $s$. To separate the responsibility, we did one step further simulation without execute the actions, getting the next state $s'$ and the attacker's action $a'$. Since the simulator is deterministic, the reward function can be further extended to $r(s,a,s',a')$. If at state $s'$, the victim crash, if $s$ as the pre-crash state and use responsible car's next action (included in $s'$) as an indicator for MEE. The pre-crash situation and reward is defined in Table \ref{table:attackerReward}. In this work, instead of implementing the whole RSS model, we only consider one pre-crash state. Therefore, no blame time concept as in \cite{Shashua2018} is implemented in this application. In the future, if one wants to implement the entire RSS model, we recommend modeling the attacker's policy with recurrent neural network and design the reward function accordingly.

The MEE of the responsible car is also implemented in the reward. When collision happens, we expect the responsible car to try to avoid the crash. As only one time step is being considered and the action space is discrete, we define the right choice of action as MEE for each pre-crash situation. As can be seen from Table \ref{table:attackerReward}, the pre-crash situation can be categorized as no car is on the lane marker, one car is on the lane marker and both cars are on the lane markers. More detailed categorized pre-crash scenarios are shown in the first column of Table \ref{table:attackerReward}. 

\begin{figure}[h]
    \centering
    \includegraphics[width=0.3\textwidth]{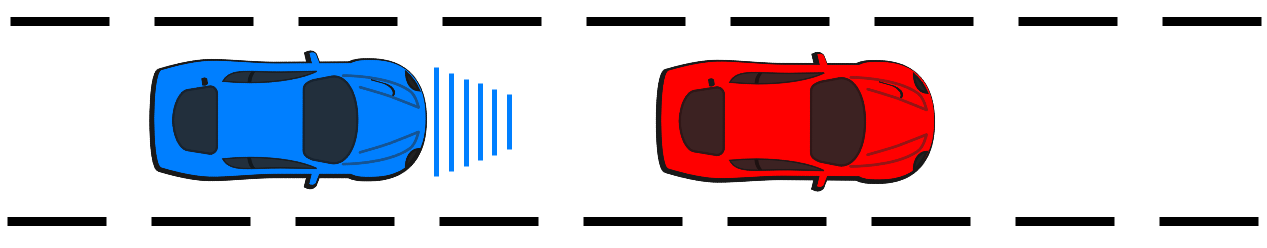}
    \caption{Pre-crash instance with no car on the lane marker}
    \label{fig:FC23}
\end{figure}

\begin{figure}[h]
    \centering
    \includegraphics[width=0.3\textwidth]{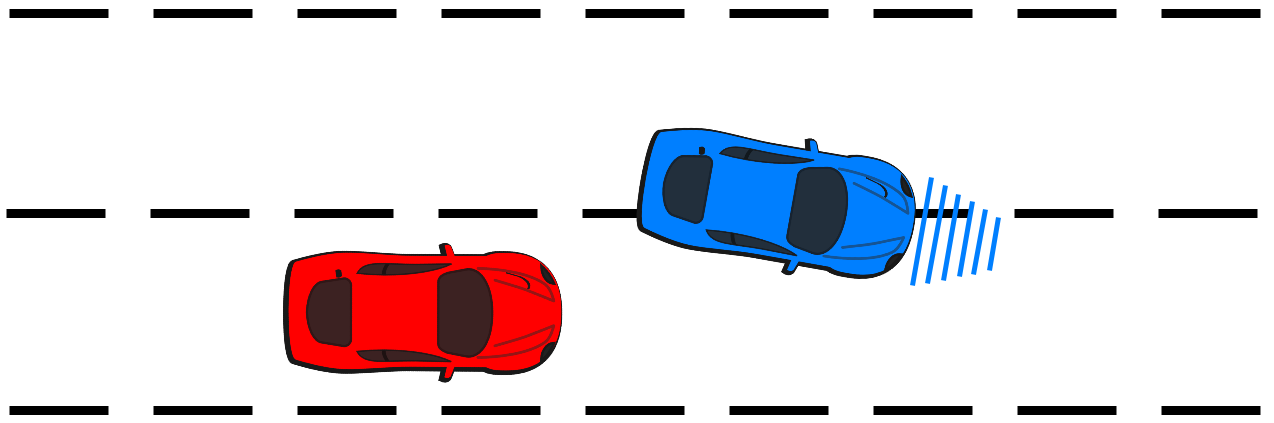}
    \caption{Pre-crash instance with only one car on the lane marker and crash with the car in the target lane}
    \label{fig:FC45}
\end{figure}

\begin{figure}[h]
    \centering
    \includegraphics[width=0.3\textwidth]{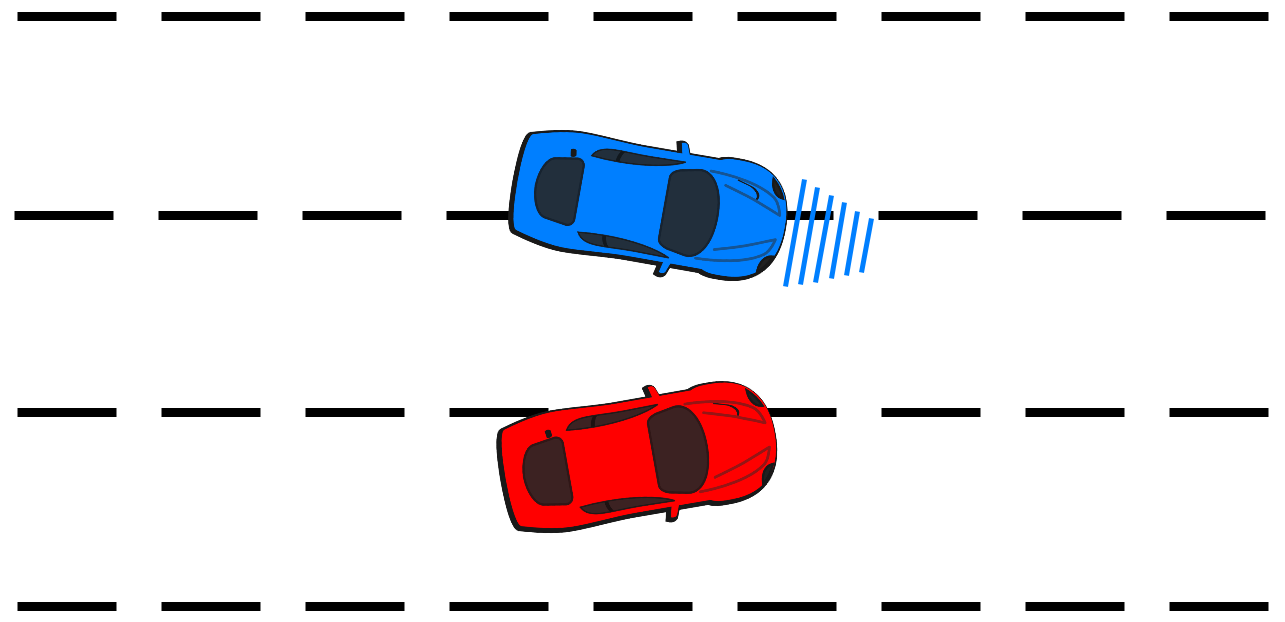}
    \caption{Pre-crash instance with both cars are on the different lane markers}
    \label{fig:FC67}
\end{figure}

As shown in Fig. \ref{fig:FC23}, when no car is on the lane marker, the rear car (blue one) is the responsible car for the crash. Therefore, the rear car is expected to conduct a hard brake to avoid the crash. Fig. \ref{fig:FC45} shows the pre-crash situation in which only one car is on the lane marker and crash with the car in the target lane. In this case, the lane change car (blue one) is the responsible car and is expected to abort the lane change to avoid the crash. As shown in Fig \ref{fig:FC67}, we illustrate the pre-crash situation in which both cars are on different lane markers. We assume that vehicles travel on the right, then in this case, both cars are at fault, but the vehicle from the left lane should yield to the vehicle from the right lane according to the Chinese traffic law \cite{ChineseTrafficLaw}. Therefore the left car (blue car) is mainly responsible for the collision and is expected to abandon the lane change to avoid the crash.

According to the responsibility and the defined MEE for each pre-crash situation, we further assigned the reward. Positive reward is given to the attacker if it lures the victim to end up with a crash which is the victim's responsibility. If the attacker can find a way to further lure the victim to directly crash into another car without MEE, the attacker will be rewarded even more. On the contrary, if the attacker is responsible for the collision, negative reward is given to the attacker.

The attacker also has a time cost of $-0.05$, which encourages the attacker to cause a collision as soon as possible. An episode is terminated if either of the following happens:
\begin{itemize}
    \item The AV crashes with another car;
    \item The attacker crashes with one car and the reward for the attacker is $-1$;
    \item The AV leaves the neighborhood of the attacker, i.e. not being one of the 6 surrounding cars of the attacker and the reward is $-1$.
\end{itemize}

We also record the Failure Code (FC) for each episode, as shown in Table \ref{table:attackerReward}. The definition of each code is described as follows:
\begin{itemize}
    \item 0: The other car is responsible for the crash (w/o. MEE);
    \item 1: The other car is responsible for the crash (w. MEE);
    \item 2: The AV crashes into the front vehicle w/o. hard brake;
    \item 3: The AV crashes into the front vehicle w. hard brake;
    \item 4: The AV changes lane and causes collision w/o. trying to abandon the lane change;
    \item 5: The AV changes lane and causes collision while trying to abandon the lane change;
    \item 6: The AV changes lane from the left to the middle lane and crashes with the car changing lane from the right to the middle lane, w/o. trying to abandon the lane change;
    \item 7: The victim changes lane from the left to the middle lane and crashes with the car changing lane from the right to the middle lane, while trying to abandon the lane change.
\end{itemize}

As shown in Table \ref{table:attackerReward}, the AV-responsible crashes, in which the AV did not use MEE, are valued most by the attacker (i.e. FC: $2$, $4$ and $6$). This kind of crash is the most harmful crash. Moreover, the action chosen by the AV's policy before these crashes definitely deserve a closer look and possible revision. In summary, instead of finding a crazy attacker, we trained a ``socially acceptable'' attacker which explores the weakness of the AV and helps to improve its policy.

\section{SIMULATION SETUP}
\label{SIMULATION SETUP}
In this Section, the simulation setup for training the attacker is described. The reinforcement learning algorithm we use is DDQN. The hyper-parameters used during training is shown in Table \ref{table:hyperparameters}. To accelerate the training, we defined two replay buffers, one is for storing normal cases and the other is for storing crash cases in which the AV agent is responsible. Therefore, the reward for socially acceptable attack and the failure code can be seen as a way to prioritise all the collected samples. Use of two replay buffers is a simpler version of prioritized experience replay as discussed in \cite{schaul2015prioritized}, which has been implemented similarly in \cite{nageshrao2019autonomous}.

\begin{table}[h]
    \centering
    \caption{Hyper-parameters for the attacker training}
    \begin{tabularx}{\linewidth}{c|l|c}
         & Description & Value\\
        \hline
        $\gamma$ & Discount factor & $0.9$ \\\hline
        $\Delta t$ & Sampling time & $1$ sec \\\hline
        $\rho$ & Learning rate & $1e{-6}$\\\hline
        $\epsilon_0$ & Starting value for $\epsilon$-greedy exploration & $0.2$\\\hline
        $C$ & Annealing factor for $\epsilon$-greedy exploration & $2e{-6}$\\\hline
        $T$ & Steps for each episode & $200$\\\hline
        $E$ & Total training episode & $1e5$\\\hline
    \end{tabularx}
    \label{table:hyperparameters}
\end{table}

During the evaluation phase, the AV is evaluated in the original environment (without the attacker) and then the same environment with one attacker. The AV is evaluated in both environments, with the total number of cars being $10$, $15$ and $20$.  The AV is evaluated for $1e6$ episodes and each episode lasts for $200$ steps unless terminated early due to a crash.   

\section{RESULTS}
\label{RESULTS}

In this Section, both the training curve of the attacker and the evaluation results of the AV in different environments are reported. As described in \cite{nageshrao2019autonomous}, after training, the AV's policy become safe and stable in the original environment. 

\begin{figure}[h]
    \centering
    \includegraphics[width=0.48\textwidth]{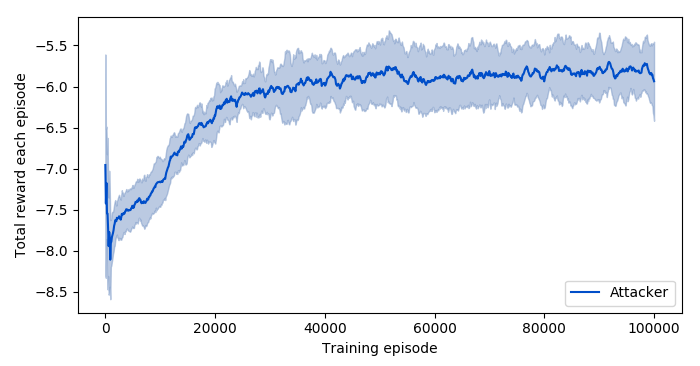}
    \caption{Average reward from 10 evaluation roll-outs with confidence bound}
    \label{fig:CurveAtt}
\end{figure}

As shown in Fig \ref{fig:CurveAtt}, the attacker is trained for $1e5$ episode and evaluated by $10$ roll-outs every 100 episodes. We train the attacker 10 times and average the evaluation roll-outs rewards. As can be seen from the Fig. \ref{fig:CurveAtt}, the attacker's policy is stable after $1e5$ training episodes. 

As described in Section \ref{SIMULATION SETUP}, the victim agent is then evaluated in two environments: the original environment and the environment with one trained attacker. In both environment, the victim is evaluated for $1e6$ episodes. The results are shown in Table \ref{table:results}. The definition of failure code is described in Section \ref{ATTACK REWARD} and here we only focus on the AV's responsible crashes (failure code 1 to 7). In the bottom of Table \ref{table:results}, we report both the number of crashes between the attacker and the victim, and the total number of crashes.  

\renewcommand{\arraystretch}{1.6}
\begin{table}[h]
    \centering
    \caption{Number of crashes during evaluation}
    \begin{tabularx}{0.99\linewidth}{cc|c|c|c|c|c|c|c}
        \multicolumn{2}{c|}{\multirow{2}{*}{\# of env. cars}} & \multicolumn{7}{c}{Failure Code}\\\cline{3-9}
        & & 1 & 2 & 3 & 4 & 5 & 6 & 7 \\\hline
        \multicolumn{1}{c|}{\multirow{3}{*}{\parbox{1cm}{w/o.\\ attacker}}} & \parbox{0.4cm}{10} & 0 & 0 & 5 & 0 & 46 & 0 & 0 \\\cline{2-9}
        \multicolumn{1}{c|}{} & \parbox{0.4cm}{15} & 0 & 0 & 14 & 0 & 55 & 0 & 0 \\\cline{2-9}
        \multicolumn{1}{c|}{} & \parbox{0.4cm}{20} & 0 & 0 & 12 & 0 & 33 & 0 & 0\\\hline
        \hline

        \multicolumn{1}{c|}{\multirow{3}{*}{\parbox{1cm}{w. one \\ attacker}}} & \parbox{0.4cm}{10} & \parbox{0.5cm}{656/\\657} & \parbox{0.5cm}{239/\\239} & \parbox{0.5cm}{180/\\186} & \parbox{0.4cm}{43/\\78} & \parbox{0.5cm}{101/\\156} & \parbox{0.5cm}{504/\\504} & \parbox{0.2cm}{9/\\9} \\[+0.5em]\cline{2-9}

        \multicolumn{1}{c|}{}& \parbox{0.4cm}{15} & \parbox{0.5cm}{447/\\448} & \parbox{0.5cm}{172/\\172} & \parbox{0.5cm}{163/\\164} & \parbox{0.4cm}{26/\\45} & \parbox{0.5cm}{63/\\88} & \parbox{0.5cm}{283/\\283} & \parbox{0.2cm}{3/\\3} \\[+0.5em]\cline{2-9}

        \multicolumn{1}{c|}{}& \parbox{0.4cm}{20} & \parbox{0.5cm}{419/\\598} & \parbox{0.5cm}{168/\\168} & \parbox{0.5cm}{137/\\139} & \parbox{0.4cm}{20/\\32} & \parbox{0.5cm}{59/\\64} & \parbox{0.5cm}{237/\\273} & \parbox{0.2cm}{1/\\1} \\[+0.5em]\hline
    \end{tabularx}
    \label{table:results}
\end{table}

As can be seen from the table, there are many more crashes when one attacker is introduced. For crashes with failure code 2, 4 and 6, where the victim is mainly responsible, the number of crashes jump from $0$ to hundreds. This is a proof that our proposed method can modify the environment transition model to increase risky but socially acceptable behaviors from other vehicles, which can provide hints on how the AV policy can be further improved.

\begin{figure}[h]
    \centering
    \includegraphics[width=0.48\textwidth]{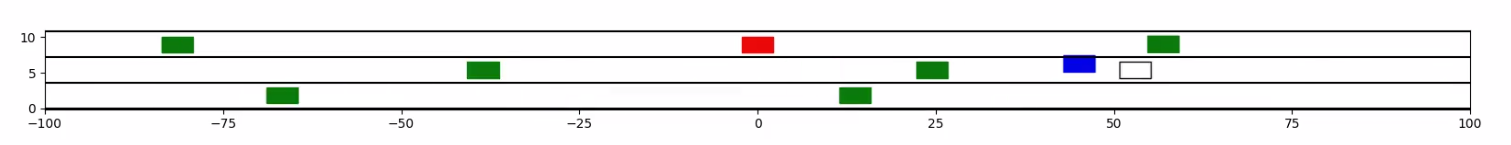}
    \caption{The victim (blue car) instead of braking and crashed by the following environment car or changing lane, it actually abandons the lane change and crashes into the front car}
    \label{fig:rewardFault}
\end{figure}

One reason why our method works, is that the original reward function for training the AV does not consider the crash responsibility. As can be seen from Fig \ref{fig:rewardFault}, the victim (blue car) is facing a situation where it will either crash into the front vehicle or crashed by the rear vehicle. Since the two situations are valued the same by the victim, in this episode, the victim actually crashes into the front vehicle.

\section{CONCLUSIONS}
\label{CONCLUSIONS}

This paper shows that the AV policy learned by deep reinforcement learning can be fragile, i.e., can still result in accidents and collisions even when the vehicles around it behave in a socially acceptable fashion. We design an attack agent by perturbing the environment transition model to induce collisions that AV would be at fault. In this work, one attacker agent is added to the environment, and the safety of the AV drops significantly. The identified collision cases can also be used to train a new AV policy.


There is a lot of potential to extend the work presented in this paper. In this paper, we apply this socially acceptable attack in an environment with discrete action space. However, it can be extended to environment with continuous action space by modeling the MEE with continuous action.  Although we add only one attacker to the environment, it can be extended to multi-attacker scenarios. Finally, one can also extend our approach by designing a Markov Game and train both the victim and the attacker together. 

\addtolength{\textheight}{-12cm}   




\section*{ACKNOWLEDGMENT}

We would like to thank the authors in paper \cite{nageshrao2019autonomous} for making the our AV policy implementation possible. 

\begin{CJK*}{UTF8}{gbsn}
\bibliography{main.bib}{}
\bibliographystyle{ieeetr}
\end{CJK*}
\end{document}